\documentclass{egpubl}
\usepackage{eurovis2023}
\usepackage[subtle]{savetrees}

\usepackage[utf8]{inputenc}
\usepackage[T1]{fontenc}
\usepackage{dfadobe}  

\usepackage{cite}  

\BibtexOrBiblatex
\electronicVersion
\PrintedOrElectronic
\ifpdf \usepackage[pdftex]{graphicx} \pdfcompresslevel=9
\else \usepackage[dvips]{graphicx} \fi

\usepackage{egweblnk}
\usepackage[fencedCode]{markdown}
\usepackage{fancyvrb}
\usepackage{relsize}
\usepackage[labelfont=bf]{caption}
\usepackage{subcaption}
\usepackage{float}

\usepackage{etoolbox}
\newcommand{\iflabelexists}[3]{\ifcsundef{r@#1}{#3}{#2}}

\usepackage{spverbatim}

\newcommand{\osfurl}{https://osf.io/n5rxd/?view_only=f09cbc2621f44074810b7d843f1e12f9}

\title[Graph Layout Algorithms in ChatGPT]%
      {Ask and You Shall Receive (a Graph Drawing): Testing ChatGPT's Potential to Apply Graph Layout Algorithms}

\author[Sara Di Bartolomeo, Giorgio Severi, Victor Schetinger, Cody Dunne]
{\parbox{\textwidth}{\centering Sara Di Bartolomeo$^1$ \orcid{0000-0001-9517-3526}, Giorgio Severi$^1$ \orcid{0000-0002-0031-2683}, Victor Schetinger$^2$ \orcid{0000-0002-8116-794X}, Cody Dunne$^1$ \orcid{0000-0002-1609-9776} \\ $^1$ Northeastern University, $^2$ TU Wien} \vspace{-0.3cm}} 

\usepackage{hyperref}
\usepackage[dvipsnames]{xcolor}
\usepackage{amsmath}
\usepackage{cleveref}
\usepackage{listings}
\usepackage{url}

\usepackage{breakurl}
\usepackage{inconsolata}
\usepackage[most]{tcolorbox}
\tcbuselibrary{listings}
\hypersetup{breaklinks=true}
\usepackage{fancyvrb}
\usepackage{fvextra}

\definecolor{codegreen}{rgb}{0,0.6,0}
\definecolor{codegray}{rgb}{0.5,0.5,0.5}
\definecolor{codepurple}{rgb}{0.58,0,0.82}
\definecolor{backcolor}{HTML}{E5E4E2}
\definecolor{altbackcolor}{HTML}{DCDCDC}

\lstdefinestyle{mystyle}{
    numbers=none,
    basicstyle=\fontsize{7}{7}\ttfamily,
    breakatwhitespace=true,         
    breaklines=true,                 
    captionpos=b,                    
    showspaces=false,                
    showstringspaces=true,
    showtabs=false,
    tabsize=1,
    breakindent=0pt,
}

\newenvironment{gptquery}{%
  \tcblisting{listing only,colback=backcolor,colframe=backcolor,top=-2pt,bottom=0pt, left=2pt, right=2pt, enhanced, breakable,
  overlay={\node[draw,fill= black,yshift=4pt,xshift=-10pt,left,text=white,
         anchor=east,font=\footnotesize\ttfamily] at (frame.south east)
         {Query};}, 
  listing options={basicstyle=\fontsize{6.5}{6.5}\ttfamily,breaklines=true,
    language=HTML},}}
{\endtcblisting}

\newenvironment{gptanswer}{%
  \tcblisting{listing only,colback=altbackcolor,colframe=altbackcolor,top=-2pt,bottom=0pt, left=2pt, right=2pt, enhanced, breakable,
  overlay={\node[draw,fill= black,yshift=4pt,xshift=-10pt,left,text=white,
         anchor=east,font=\footnotesize\ttfamily] at (frame.south east)
         {Answer};}, 
  listing options={basicstyle=\fontsize{6.5}{6.5}\ttfamily,breaklines=true,
    language=HTML},}}
{\endtcblisting}

\begin{document}


\maketitle
\begin{abstract}
Large language models (LLMs) have recently taken the world by storm. 
They can generate coherent text, hold meaningful conversations, and be taught concepts and basic sets of instructions---such as the steps of an algorithm. 
In this context, we are interested in exploring the application of LLMs to graph drawing algorithms by performing experiments on ChatGPT.
These algorithms are used to improve the readability of graph visualizations. 
The probabilistic  nature of LLMs presents challenges to implementing algorithms correctly, but we believe that LLMs' ability to learn from vast amounts of data and apply complex operations
may lead to interesting graph drawing results. 
For example, we could enable users with limited coding backgrounds to use simple natural language to create effective graph visualizations.
Natural language specification would make data visualization more accessible and user-friendly for a wider range of users.
Exploring LLMs' capabilities for graph drawing can also help us better understand how to formulate complex algorithms for LLMs; a type of knowledge that could transfer to other areas of computer science.
Overall, our goal is to shed light on the exciting possibilities of using LLMs for graph drawing while providing a balanced assessment of the challenges and opportunities they present. 
A free copy of this paper with all supplemental materials required to reproduce our results is available on \href{\osfurl}{osf.io (anonymous link)}.


\end{abstract}
\section{Introduction}

\begin{figure*}[tb]
\centering
     \begin{subfigure}[t]{0.49\textwidth}
         \centering
         \includegraphics[width=\linewidth]{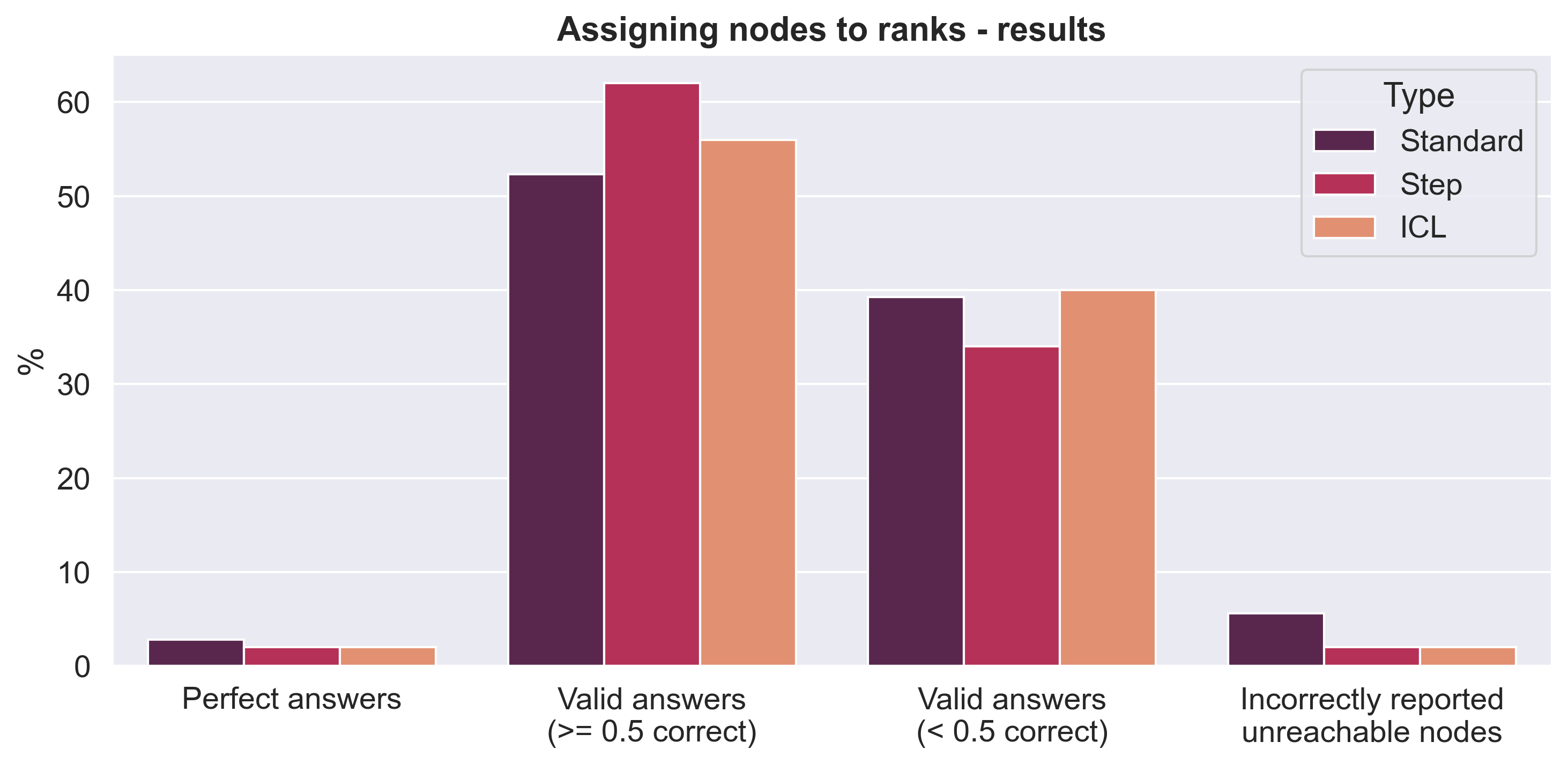}
         \caption{%
         We asked ChatGPT to assign each node to a layer (rank) based on its shortest path to a source.
         Few rank assignments were perfect (i.e.\ they obtained the minimum edge length using a correct breadth-first search), but we were pleased that the bulk were at least valid and that most had at least half the nodes correctly assigned.}
        \label{fig:rank_assignment_results2}
     \end{subfigure}
     \hfill
     \begin{subfigure}[t]{0.49\textwidth}
         \centering
         \includegraphics[width=\linewidth]{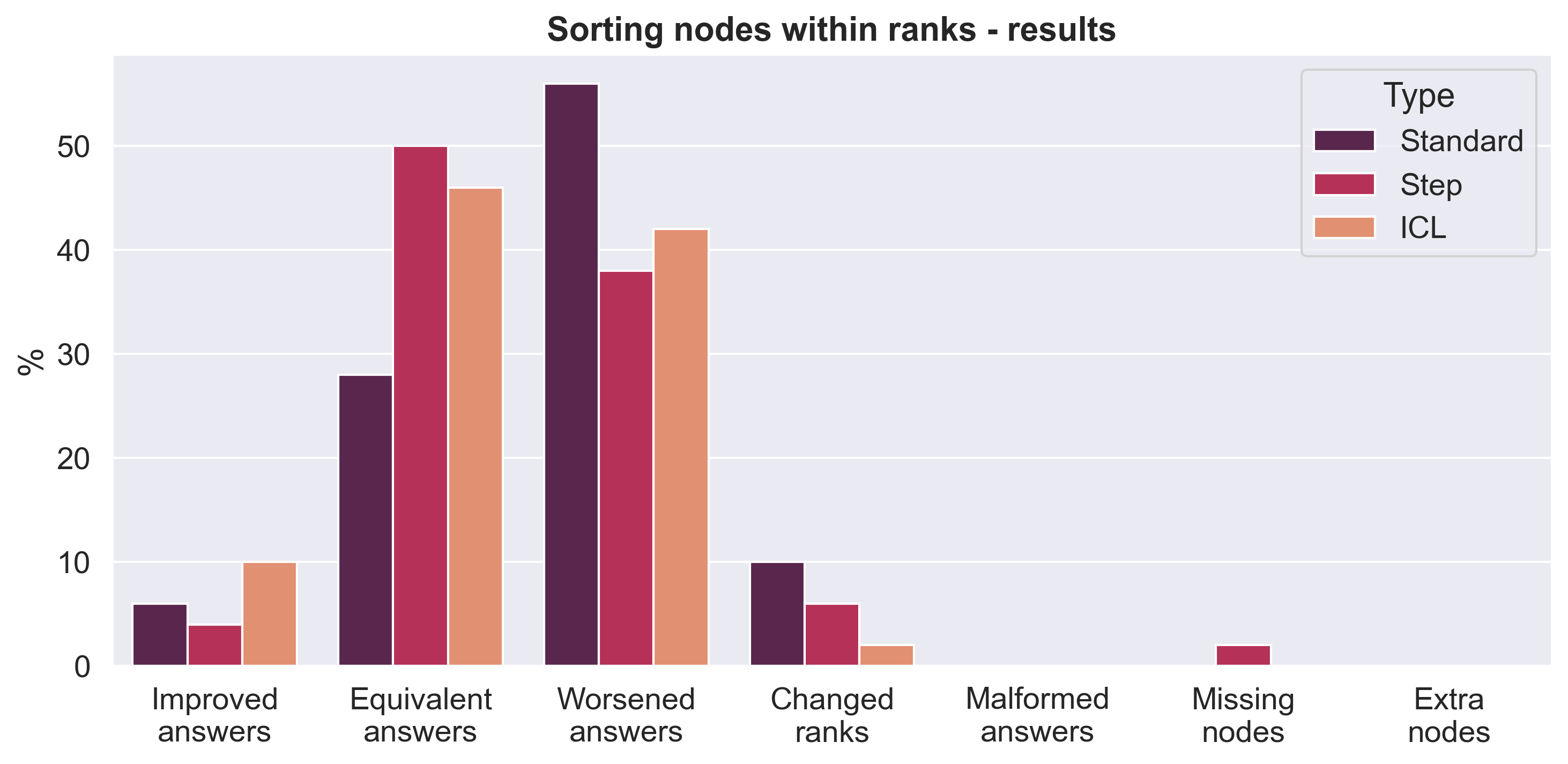}
        \caption{%
        We asked ChatGPT to sort nodes in each layer according to the median position of their neighbors in the next layer.
        We compared the answer to the number of crossings in the input graph. 
        Using the \textbf{Step} and \textbf{ICL} approach produced equivalent results the majority of the times.
        \textbf{Standard} returned consistently worse results. }
        \label{fig:transpose_results}
     \end{subfigure}
     \hfill
     \begin{subfigure}[t]{0.49\textwidth}
         \centering
         \includegraphics[width=\linewidth]{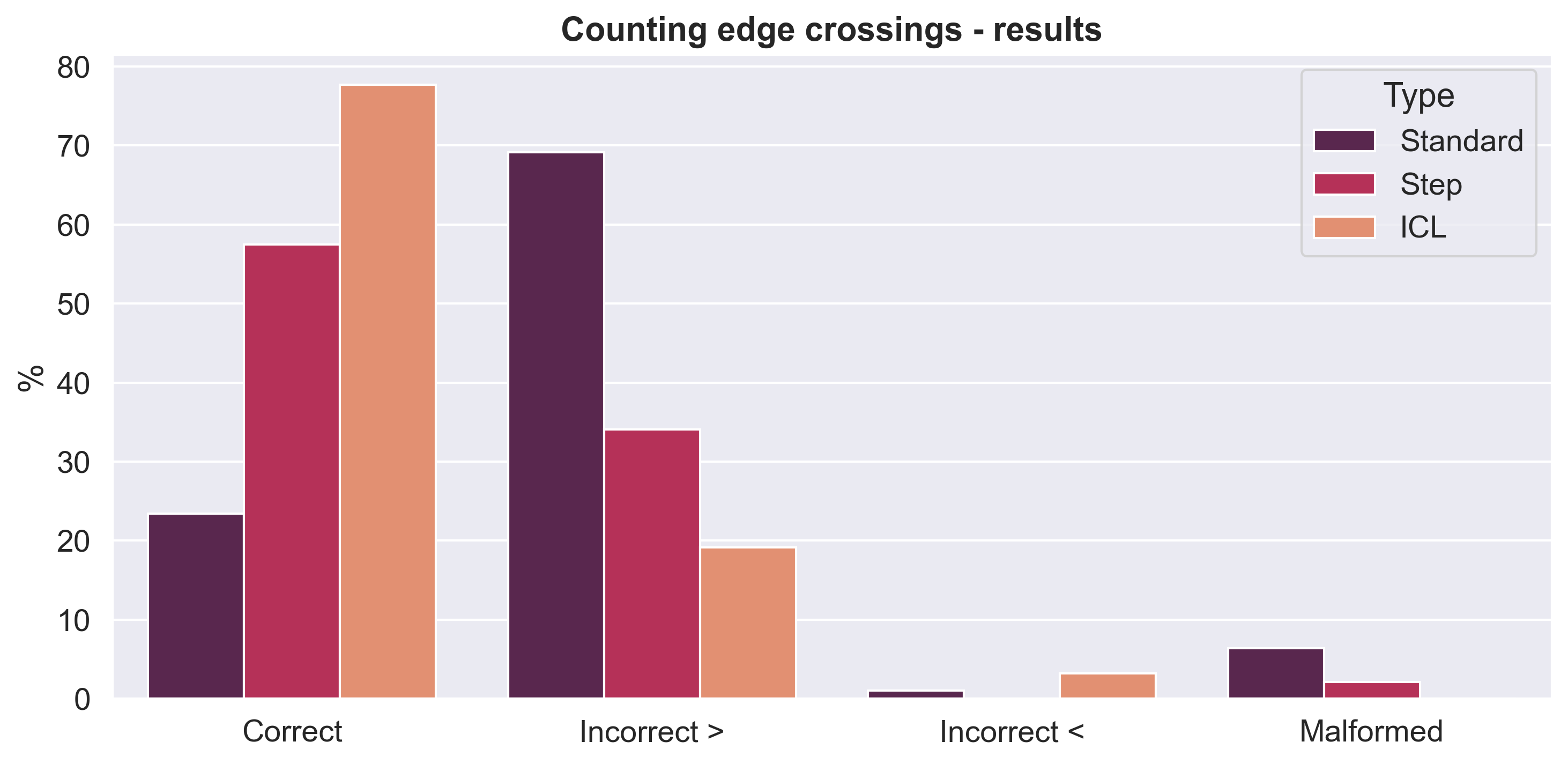}
        \caption{%
        When asking ChatGPT to count crossings on bipartiate graphs, we found that \textbf{ICL} worked better than reasoning \textbf{Steps}, which in turn outperformed \textbf{Standard} prompts. 
        ``Incorrect >'' means that ChatGPT over-estimated the number of crossings, while ``Incorrect <'' were underestimates.
        A few answers were malformed.}
        \label{fig:crossing_results}
     \end{subfigure}
     \hfill
     \begin{subfigure}[t]{0.49\textwidth}
         \centering
         \includegraphics[width=\linewidth]{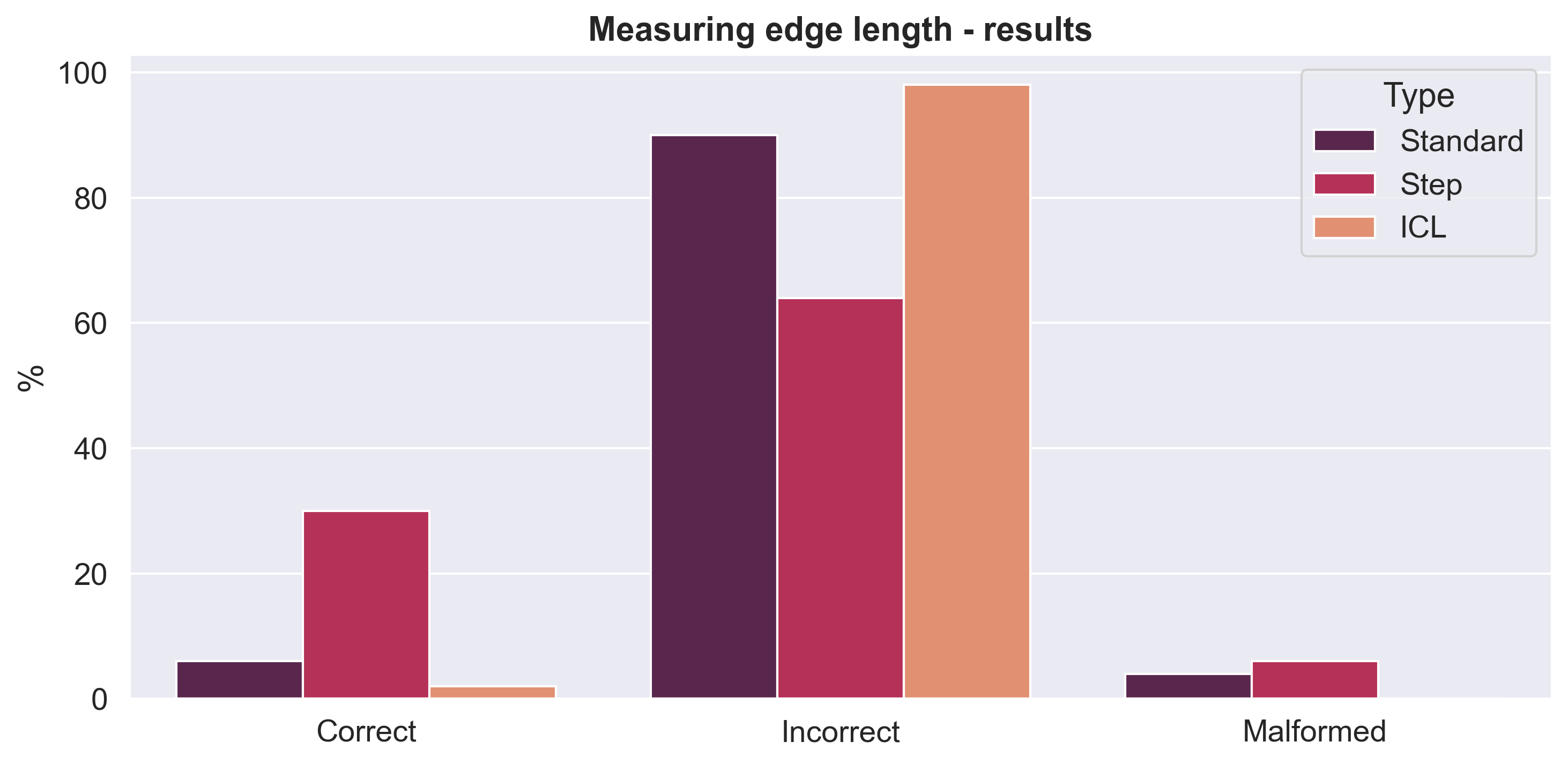}
        \caption{%
        Measuring edge length with ChatGPT worked best with reasoning \textbf{Steps}, with 25\% answered correctly.
        The incorrect results had an average error of 3.69.
        \textbf{ICL} was only correct in 2\% of cases (average error 7.88), but never returned malformed answers.
        Low correctness may be due to the LLM being prone repeat example values.}
        \label{fig:count_edge_length}
     \end{subfigure}
        \caption{%
            Results from our experiments asking ChatGPT to perform graph drawing tasks.
            See more results in our
            \iflabelexists{app:task-format-conversion}
                {appendices}
                {appendices at \href{\osfurl}{osf.io (anonymous link)}}.
            }
        \label{fig:three graphs}
        \vspace{-0.4cm}
\end{figure*}

\vspace{-0.2cm}
A graph layout algorithm maps nodes and edges in a graph to coordinates in space---an essential step in rendering visible its abstract topology.
These algorithms generally optimize for readability criteria such as reducing the number of edge crossings in the resulting drawing \cite{PURCHASE2002501, di_bartolomeo_crnovrsanin_saffo_dunne_2023}.
Decades of research in the field have produced many layout algorithms, such as the popular Sugiyama algorithm \cite{sugiyama1981} for layered graphs, which we use in this paper.
When considering the needs of a user, however, it is challenging to understand which algorithm to choose and how to control its parameters to obtain a desired result \cite{kwonDeepGenerativeModel2020}.
Whether the graph is a social network, -omics diagrams, or a subway map, a domain expert will have an implicit, subjective expectation of what needs to be seen.
Translating these needs into a choice of aesthetic criteria is neither a simple nor an exact task.


An ideal system would let a user provide a graph and explain in their own words how to visualize it.
For example, an Art Historian could express their needs as ``I want to see the collaboration network of Kandinsky with him at the center, thicker edges showing more co-exhibitions, and with all other Russian painters visible''.
In contrast, a graph drawing researcher might say: ``I want to minimize edge crossings and...'' 
This interaction is not yet possible, but OpenAI's ChatGPT has recently enabled the general public to use large language models (LLMs) and demonstrates the potential for natural language interfaces.


In September 2022, Jacob Brazeal \cite{gptpathfinding} described an experiment using GPT-3 to run a path-finding algorithm.
His results showed the model could apply multiple steps correctly, inspiring us to explore the feasibility of using ChatGPT for more complicated graph layout algorithms.
In particular, we dissected a layout algorithm into multiple ``bite-sized'' tasks that a language model could interpret.
Since both inputs and outputs of these algorithms can be represented as text, e.g., inputting a list of nodes and edges and outputting a table of coordinates, in theory LLMs could act as general-purpose solvers. 
Applying a graph layout algorithm via a generative text model would require no programming, just a natural-language description of the problem and the graph.
This may enable users to more rapidly specify novel constraints on the layout.
However, generative models have downsides: (a) they require careful consideration of the words used to describe the graph and the problem, and (b) due to their stochastic nature, the correctness of the outputs cannot be ensured a priori.

To explore the possible benefits and downsides of this approach, we designed a set of experiments that would help illustrate the art of the possible as well as evaluate LLM correctness against existing algorithms.
We also discuss how to best formulate graph drawing problems so that they are easily understood by an LLM.
Even though some mistakes and imprecision in the returned solutions are to be expected using this heuristic LLM-based layout algorithm, our results show that the majority of the results returned are valid solutions.
These promising results, combined with the fast pace at which LLMs are evolving, lead us to believe that we can expect more effective layout results in the near future.
All the code used for the analysis---as well as our supplemental material with experiments, all queries, and all answers---is available on \href{\osfurl}{osf.io (anonymous link)}.

\section{Background} 

A language model defines a probability distribution over a sequence of linguistic units, and can be used to predict the most likely next unit in a sequence.
ChatGPT~\cite{openaiChatGPTOptimizingLanguage2022}, a recent addition to the GPT (Generative Pre-trained Transformer) family~\cite{radfordImprovingLanguageUnderstanding2018, radfordLanguageModelsAre2020, brownLanguageModelsAre2020a} of causal language models, is currently regarded as the state of the art of conversational agents.
Given an existing sequence of tokens, where each token represents a pre-defined sub-component of a natural language word, corresponding to a unique integer number, these models are trained to iteratively and autoregressively predict the most likely next tokens.
These tokens, chained-together, end up forming words, sentences, and even entire documents.
This characteristic training procedure distinguishes them from the other famous family of transformer models (represented by BERT~\cite{devlinBERTPretrainingDeep2019}) trained for masked language modeling, where the objective is to learn to fill in missing tokens in a sequence.

In contrast to its predecessors, ChatGPT leverages Reinforcement Learning from Human Feedback (RLHF) to align responses generated by the model with the expectations of end users.
The exact details of ChatGPT's model have not been fully disclosed by OpenAI.
The closest documented system is \textsc{text-davinci-003}~\cite{openaiOpenAIAPI2023}, which is described as a version of InstructGPT~\cite{ouyangTrainingLanguageModels2022} fine-tuned with RLHF.
We are not in a position to fine-tune ChatGPT for our tasks, but we believe that its RLHF training, together with its simple interface by which users can specify complex prompts, makes it particularly well-suited to follow the user-specified steps of a graph layout algorithm.

Beyond \emph{natural} language, GPT models have shown impressive performance when generating different textual data such as programming language code~\cite{chenEvaluatingLargeLanguage2021}, and multiple \emph{emergent} abilities~\cite{weiEmergentAbilitiesLarge2023} have been observed with the progressive increases in model capacity and training volumes.
We can situate our paper within a larger body of work that tries to explore such emergent abilities.
The majority of this exploration has been focused in NLP tasks~\cite{koconChatGPTJackAll2023,mahowaldDissociatingLanguageThought2023}, which is not strictly our case.
However, the manipulation of a graph's topological space within the internal representation of a LLM can be related to the problem of grounded conceptual spaces \cite{patelMappingLanguageModels2022}.
The fast evolution of generative models has brought growing interest in using them for data visualization \cite{schetinger2023}.
Although neural network approaches have been tested for graph layout algorithms \cite{Giovannangeli2021, klammler2018, felice2019}, this iteration of generative models is so recent that there has been little research yet on applying them to graph drawing.
Hence, it is our intention to test to what extent these models can be used to apply graph layout algorithms.

\section{Experiments}

We re-create the classic Sugiyama \cite{sugiyama1981} layout algorithm for layered graphs using ChatGPT.
This algorithm had several component tasks we could test, as well as several utility tasks for us to evaluate.
It is important to keep in mind that the expected result of a graph layout algorithm is a coordinate assignment, mapping nodes to coordinates in space. 
Thus, we expected the LLM to generate numerical values for every node, \textit{not} a graphical rendering.
While we do experiment with generating SVG illustrations
\iflabelexists{app:drawing}
    {(\cref{app:drawing})}
    {in our online appendices (\href{\osfurl}{osf.io (anonymous link)})},
rendering is not the focus of this paper.
We treat rendering as a successive step which can be done with any graphical library.

The Sugiyama algorithm incorporates several steps: (1) cycle removal, (2) layer assignment, (3) sorting nodes within layers, and (4) final positioning.
Here, we prioritize discussing the pivotal steps of layer assignment (\cref{sec:task-rank-assignment}) and sorting nodes within layers (\cref{sec:task-sorting}).
We also explore the related tasks of counting crossings and edge length to evaluate the quality of the layout (\cref{sec:task-count-cross-edge-length}).
Actual queries to ChatGPT and additional discussion of these tasks is available in  
\iflabelexists{app:task-format-conversion}{\cref{app:task-layer-assignment,app:task-sorting-nodes,app:task-count-crossings,app:task-edge-length}}
    {our online appendices at \href{\osfurl}{osf.io (anonymous link)}}.
Our appendices also detail additional experiments with executing several utility (and fun!) tasks using ChatGPT.
These include 
explaining the Sugiyama algorithm in poetry\iflabelexists{app:task-poetry}{ (\cref{app:task-poetry})}{},
generating test graphs\iflabelexists{app:task-graph-generation}{ (\cref{app:task-graph-generation})}{},
converting between file formats\iflabelexists{app:task-format-conversion}{ (\cref{app:task-format-conversion})}{},
defining additional graph properties\iflabelexists{app:task-defining-properties}{ (\cref{app:task-defining-properties})}{},
generating graphs from scenes (e.g.\ for movie StoryLines \cite{Tanahashi_designconsiderations})\iflabelexists{app:generate-scene-from-graph}{ (\cref{app:generate-graph-from-scene})}{}
as well as entirely new scenes from graphs\iflabelexists{app:generate-scene-from-graph}{ (\cref{app:generate-scene-from-graph})}{},
and creating SVG illustrations of the graph\iflabelexists{app:drawing}{ (\cref{app:drawing})}{}.

We ran our experiments using graphs from Rome-Lib \cite{DIBATTISTA2000}, a popular benchmark dataset for graph layout algorithms. 
We used only the graphs with 10 or 11 nodes so that we did not exceed the fixed budget of tokens that ChatGPT could process simultaneously.
To test the correctness of the results, every answer from ChatGPT was compared against a ground truth; we show the the results from these comparisons in our figures.
All the examples presented were ran on clean chat threads to avoid previous inputs contaminating the results.
We used OpenAI's web interface (at the time of writing there is no available API) and the ChatGPT Plus default model from 2023-02-13 -- 2023-03-01.
In order to run multiple tests in sequence, we developed a script that takes graph problems as input and simulates a user typing the query into the web interface.

Although it is possible to ask ChatGPT to write code to execute the tasks we defined, we wanted to test its ability to apply the algorithm and reason on the problems without executing any code.
We tried several approaches to formulating problems.
In some cases, we gave ChatGPT examples of solved problems in the query along with the usual explanation of the task.
We mark this In-Context Learning (\textbf{ICL})~\cite{xieExplanationIncontextLearning2022} when it is used.
To construct these examples, we randomly sample $k=[3, 5]$ other instances of the task from our pool, and augment the prompt with the input and correct answers to those instances.
We also experimented with splitting the task in to \textit{reasoning steps} and asking ChatGPT to provide answers for each step (these are marked \textbf{Step}).
This technique is also known as chain-of-thought prompting (CoT).
We use a zero-shot CoT~\cite{kojimaLargeLanguageModels2022}, which means that we do not provide step-by-step examples of the solution to similar instances of the task, but rather ask the model to write down the detailed explanation for each step towards the original task.
Cases in which we explained the task to ChatGPT but provided no examples and requested no reasoning steps are marked as \textbf{Standard}.
Examples and comparisons between the different approaches can be found in the appendix.

\subsection{Layer assignment} \label{sec:task-rank-assignment}

Layer assignment (a.k.a.\ rank assignment) is an important step in the Sugiyama layout \cite{sugiyama1981}, as well as the default \textit{dot} algorithm \cite{gansner1993technique} in Graphviz. 
Graphs without an inherent layering must have each node assigned to a layer before a layered graph layout algorithm can be used. 
Of the many methods for assigning layers \cite{sugiyama2002,
Healy2013HierarchicalDrawing}, we choose a simple approach: select a source, then assign to every node a layer that is equal to the length of the shortest path from it to the source.
I.e., a node that is 2 hops from a source on layer 0 will be assigned to layer 2. 
This is the method used by \textsc{Stratisfimal Layout} \cite{dibartolomeo2021stratisfimal}.
(Note that layer assignment with directed graphs is usually preceded by a cycle removal step; we considered all edges as undirected.)
This straightforward approach requires either recursion or a queue data structure to conduct a breadth-first search.


We provide the example queries and answers in 
\iflabelexists{app:task-layer-assignment}
    {\cref{app:task-layer-assignment}}
    {our online appendices at \href{\osfurl}{osf.io (anonymous link)}}.
For each graph, we recorded the percentage of nodes assigned to the correct layer.
\Cref{fig:rank_assignment_results2} shows us that ChatGPT rarely assigns the layers perfectly and occasionally will even incorrectly report nodes as unreachable. 
However, the vast bulk of answers were at least valid assignments and most of the time ChatGPT assigned at least half of the nodes correctly.
Interestingly, ChatGPT often recognized in its answers that what we were asking was the application of a breadth-first search---even without us specifying so in the prompt.

\begin{figure}[t]
    \centering
    \includegraphics[width=.9\linewidth]{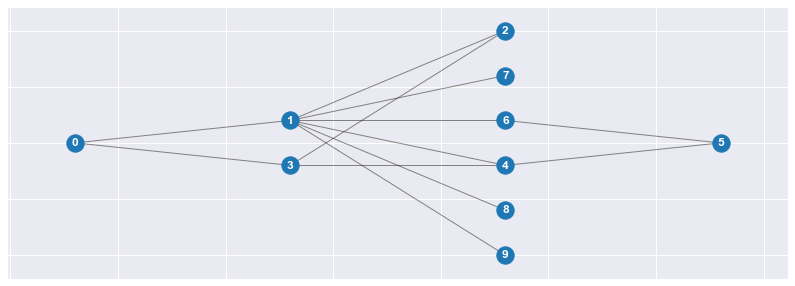}
    \includegraphics[width=.9\linewidth]{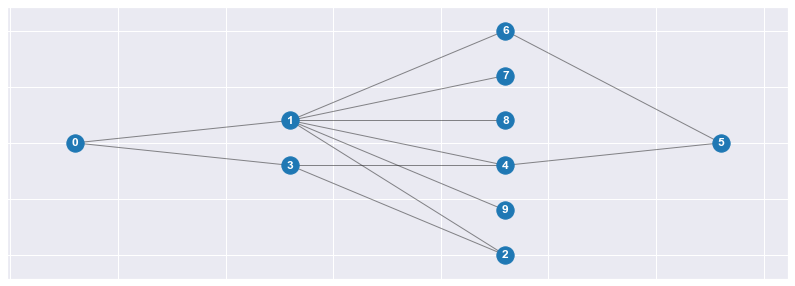}
    \caption{An example of a more readable graph produced by asking ChatGPT to sort nodes in each layer using the median heuristic.
    The input graph (top) has 7 crossings, while the output (bottom) has only 2.}
    \label{fig:graph_rendering}
    \vspace{-0.3cm}
\end{figure}


\subsection{Sorting nodes within layers}
\label{sec:task-sorting}

The next step of the Sugiyama algorithm \cite{sugiyama1981} is to sort nodes within each layer.
There are multiple approaches (see \cite{sugiyama2002,
Healy2013HierarchicalDrawing}) but we use the \textit{median heuristic}.
We sequentially sweep across the layers of the graph, consider each layer in turn, and position each node within it according to the median of its neighbors.
Iteratively sweeping one direction then the reverse gives us a relative positioning of the nodes within each layer, from which they can be evenly distributed (e.g.\ on a grid).
Thus, in a horizontal layered graph, the $x$ position of a node is defined by its layer, while the $y$ position is defined by this sorting. 
While the process is meant to be repeated for many iterations or until convergence, we asked ChatGPT to do it only once.
To simplify the description in the following steps, we removed edges that connect nodes that share the same layer.
The results are illustrated in \Cref{fig:transpose_results}.
\Cref{fig:graph_rendering} shows a rendering of a graph given as input vs.\ a rendering of ChatGPT's response.
Below is an example prompt used for this task, followed by the corresponding answer. 
See more examples in
\iflabelexists{app:task-sorting-nodes}{\cref{app:task-sorting-nodes}}{our online appendices at \href{\osfurl}{osf.io (anonymous link)}}.

\begin{gptquery}
This is a list of directed edges in a layered graph.
[(0, 1), (1, 2), (3, 2), (0, 3), (4, 5), (6, 5), (1, 7), (1, 6), (1, 4), (3, 4), (1, 8), (1, 9)]

This is the description of what nodes are contained 
in what layer:  
Layer 0: [0] 
Layer 1: [1, 3] 
Layer 2: [2, 7, 6, 4, 8, 9] 
Layer 3: [5] 

Write no code and no explanations.

Start from layer 0. 
Consider every node in layer 0, and for each node consider 
its neighbors in the next layer. 
For each node in layer 0, calculate the positions 
of the neighbors in the next layer. 
Then, sort the nodes in layer 0 based on the median 
position of their neighbors.
If a number is NaN, then it is equal to 0.
If a node has no neighbors, assign an associated median 
of 0.

Repeat the process for every layer.

Finally, I want you to write the layers and the nodes in 
them once more, but the nodes in the layers have to be 
sorted according to their associated median value.
Like this:
Layer 0: [<sorted nodes in layer 0>]
If node A has an associated median value that is less 
than the associated median value of node B, then node 
A must come before node B. 
\end{gptquery}
\begin{gptanswer}
Layer 0: [0]
Layer 1: [3, 1]
Layer 2: [7, 6, 8, 9, 2, 4]
Layer 3: [5]
\end{gptanswer}


\subsection{Counting crossings or edge length}
\label{sec:task-count-cross-edge-length}
\textbf{Counting crossings} is a fundamental step to evaluate the quality of a layout.
With our current graph representations, crossings can only appear between one layer and the next.
To simplify the task for the language model, we decided to split the problem in sub-graphs: each pair of consecutive layers was described as a bipartite graph.
We therefore generated $n-1$ queries for each graph, where $n$ is the number of layers in a graph.
We then compared the solutions given by ChatGPT against the ground truth.
We found that both prompt shaping techniques were drastically more effective than the \textbf{Standard} prompts in eliciting the correct answer, as shown by \Cref{fig:crossing_results}.
While \textbf{ICL} appeared to lead to the most successful outcomes, this result may be biased by the relatively large quantity of results with 0 crossings, which were thus more likely to appear in the provided examples.

\textbf{Edge length} is another key readability criteria.
The shorter the edge, the easier it is for a human reader to follow \cite{PURCHASE2002501}.
It is therefore desirable to minimize overall edge length in the drawing. 
Assuming a unitary distance between adjacent layers, we can compute the length of each edge as the absolute value of the index of the layer of the source of the edge minus the index of the target of the edge:
$|layers.indexOf(e.source) - layers.indexOf(e.target)|$.
The method we used to assign layers (\cref{sec:task-rank-assignment}), when correctly performed, produces the minimum possible edge length---each has a length of 1.
In this case, counting the total edge length is equivalent to counting the number of edges.
However, we did not explicitly provide this information to ChatGPT---instead specifying that the distance between consecutive layers was 1---and asked it to count the total edge length
On this task, ChatGPT was surprisingly able to return an exact result for every graph, without necessitating \textbf{ICL} or reasoning \textbf{Steps}.

To test this ability on a more complex case, we created a new layer assignment for 50 graphs by assigning each node to a random layer.
Thus the length of each edge was no longer always 1.
Our results from asking ChatGPT to compute edge length with these new graphs are shown in \Cref{fig:count_edge_length}.
There were now many more incorrect answers, but the differences between the performance of the different prompting approaches become much more evident: using \textbf{Steps} we had considerably better results than \textbf{ICL} or \textbf{Standard}.



\section{Discussion and conclusions} \label{sec:discussion}

There is an incredible potential for LLMs to be used in visualization and graph drawing---getting to the point where we obtain reliable results could enable users with no coding backgrounds to create novel visualizations without having to write or execute any code. 
Currently, however, we discovered substantial limitations with this approach.
But the encouraging results we obtained on some sub-tasks, coupled with the breakneck speed of LLM improvements, leads us to be optimistic about the future utility of LLMs for graph drawing tasks. 
In our            
\iflabelexists{app:task-format-conversion}
    {appendices}
    {online appendices at \href{\osfurl}{osf.io (anonymous link)}},
 we explore other tasks that could be relevant for a natural language interface for graph drawing.
 We hope the examples we provide will lead to interesting opportunities for future research.

\textbf{Potentially invalid results:}
Asking LLMs to perform layout algorithms can potentially lead to invalid results.
The stochastic nature of LLMs and the challenges involved in parsing natural language means there can be few guarantees.
Checking the solution manually or against a traditionally-computed baseline is necessary if exact answers are needed.
LLM alignment is an active field of research and result quality is expected to improve in the near future, but we doubt that the problem can be fully solved.
However, there is one promising avenue to explore.
The model we tested was not trained specifically for our algorithmic tasks, but was still able to provide mostly valid and often good answers due to its training as a next-token predictor.
Fine-tuning language models for specific tasks is a common practice which could lead to significant improvements in our graph drawing performance.


\textbf{Prompt engineering:}
It is important to keep in mind the influence of the prompt over the resulting response that is obtained from ChatGPT.
Any difference in wording can give a different result.
We experimented with several established prompting techniques, but exploring the entire spectrum of prompt-crafting is outside of our scope.
We refer the reader to Liu et al.\ \cite{liuPretrainPromptPredict2021} for a more complete investigation of this topic.

\textbf{Scalability limitations:}
Most transformer-based language models, including ChatGPT, have a fixed budget of tokens they can process simultaneously, including both input and output.
A token is a word or part of a word that represents a unitary element of the input and output sequences.
As noted earlier, we exclusively experimented with small graphs to avoid this issue.
This token limit places a size limit on input graphs, reducing the utility of LLM graph drawing approaches.
However, as the size of language models as increased, so has their token processing capacity.
We believe scalability will become less of an issue over time.



\clearpage

\bibliographystyle{eg-alpha-doi}  
\bibliography{egbibsample, prj_graphgpt}   

\clearpage



\appendix
\noindent\textbf{Appendices}

\noindent Our idealized use of a natural language interface for graph drawing involves different levels of tasks.
The core experiments reported in the paper body focused on relatively low-level tasks, mainly due to the availability of ground truths and the feasibility of obtaining quantitative results.
However, we explored many dimensions of this application space by performing a variety of experiments.
In these appendices, we report in some of the most interesting ones.
From defining novel, quirky, properties to writing scenes from an input graph, these examples show the potential of using LLMs in various graph-related problems, and we hope provide readers with many ideas for future work.

\section{Layer assignment}
\label{app:task-layer-assignment}

Here we provide more details and results for the layer assignment task detailed in
\iflabelexists{sec:task-rank-assignment}
    {\cref{sec:task-rank-assignment} of}
    {}
the paper.

\begin{figure}[h]
    \centering
    \includegraphics[width=\linewidth]{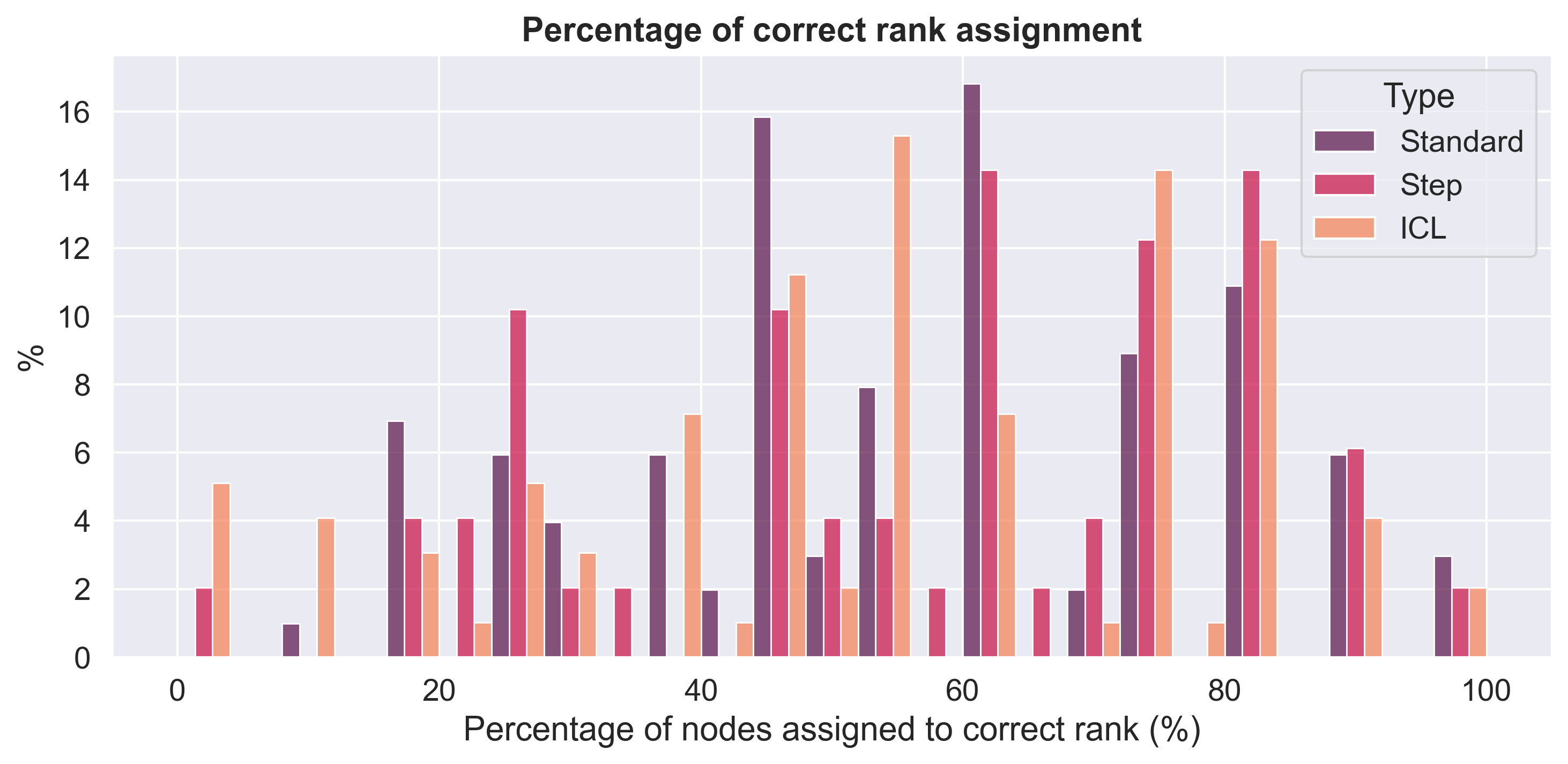}
    \caption{How many times did ChatGPT perform a correct layer assignment? The values on the x axis indicates the ratio of nodes assigned to the correct layer in a graph, while the y axis indicates how many graphs were solved with that amount of correctness. Most times, ChatGPT was able to assign more than half of the nodes to the correct layer. It is also important to note that a single node assigned to an incorrect layer can impact the assignment of all the following nodes---thus having a ripple effect on the other nodes.}
    \label{fig:rank_assignment_results}
\end{figure}

The following boxes show examples of wording we used to ask ChatGPT to perform a layer assignment (\textbf{Query}), as well as ChatGPTs answers (\textbf{Answer}).


\texttt{Layer assignment --- Standard:}
\begin{gptquery}
Graph G has 10 nodes, numbered from 0 to 9. Graph G has 10 edges.
This is the list of edge connections:
[5, 0],[6, 1],[6, 2],[2, 7],[7, 4],[8, 2],[9, 7],[9, 8],[9, 5],[3, 5]
Perform a rank assignment on the graph. Use node 0 as a source for the graph. Each node must be assigned to a rank that is equal to the shortest path between that node and the source. Thus, node 0 will be assigned to rank 0, and the neighbors of node 0 will be assigned to rank 1.
Write no explanations, only respond with the id of each node and the rank it has been assigned to in a format <id> - <rank>.
\end{gptquery}


\begin{gptanswer}
0 - 0
5 - 1
6 - 1
3 - 2
1 - 2
2 - 2
7 - 3
9 - 3
8 - 3
4 - 4
\end{gptanswer}

\texttt{Layer assignment --- Steps:}
\begin{gptquery}
You are a powerful algorithm for graph drawing. Your job is to answer questions regarding graphs. This is a list of directed edges in a graph::

edge_list: [(5, 0), (6, 1), (6, 2), (2, 7), (7, 4), (8, 2), (9, 7), (9, 8), (9, 5), (3, 5)]

Start from node 0. Node 0 belongs to layer 0.

Find all the neighbors of node 0. Assign all the neighbors of node 0 to layer number 1.
Then write the list of nodes belonging to layer 1 in a code block, one node per line, formatted as node: layer.

Find all neighbors of the nodes in layer 1 that do not already belong to a layer. Assign to them depth 2. 
Then write the list of nodes belonging to layer 2, in a code block, one node per line, formatted as node: layer.

Find all neighbors of the nodes in layer 2 that do not already belong to a layer. Assign to them depth 3. 
Then write the list of nodes belonging to layer 3 in a code block, one node per line, formatted as node: layer.

Find all neighbors of the nodes in layer 3 that do not already belong to a layer. Assign to them depth 4. 
Then write the list of nodes belonging to layer 4 in a code block, one node per line, formatted as node: layer.

Find all neighbors of the nodes in layer 4 that do not already belong to a layer. Assign to them depth 5. 
Then write the list of nodes belonging to layer 5 in a code block, one node per line, formatted as node: layer.

Repeat the process, incrementing the layer, until there are no more nodes that are not assigned to any layer
Write the final result in a code block, one node per line, formatted as node: layer.
\end{gptquery}

\begin{gptanswer}
Starting from node 0, we can perform a breadth-first search to assign a layer number to each node in the graph. Here's the algorithm step by step:

Assign node 0 to layer 0
Find all neighbors of node 0: nodes 5. Assign them to layer 1.
Find all neighbors of nodes in layer 1 that do not already belong to a layer: nodes 3, 6. Assign them to layer 2.
Find all neighbors of nodes in layer 2 that do not already belong to a layer: nodes 1, 2, 7. Assign them to layer 3.
Find all neighbors of nodes in layer 3 that do not already belong to a layer: nodes 4, 8, 9. Assign them to layer 4.
Now all nodes have been assigned to a layer, so we can write the final result:

0: 0
5: 1
3: 2
6: 1
1: 3
2: 3
7: 3
4: 4
8: 4
9: 4
\end{gptanswer}

\texttt{Layer assignment --- ICL:}
\begin{gptquery}
Perform a rank assignment on the graph. Use node 0 as a source for the graph. Each node must be assigned to a rank that is equal to the shortest path between that node and the source. Thus, node 0 will be assigned to rank 0, and the neighbors of node 0 will be assigned to rank 1. 
Write no explanations, only respond with the id of each node and the rank it has been assigned to in a format <id> - <rank>.

Input:
Graph G has 11 nodes, numbered from 0 to 10. Graph G has 15 edges.
This is the list of edge connections:
[0, 1],[0, 2],[3, 4],[4, 5],[5, 6],[6, 1],[1, 5],[3, 1],[3, 2],[1, 7],[7, 8],[8, 3],[9, 8],[10, 1],[10, 6]
Answer:
0 - 0
1 - 1
2 - 1
6 - 2
5 - 2
3 - 2
7 - 2
10 - 2
4 - 3
8 - 3
9 - 4

Input:
Graph G has 11 nodes, numbered from 0 to 10. Graph G has 12 edges.
This is the list of edge connections:
[0, 7],[7, 1],[6, 2],[2, 5],[5, 3],[4, 8],[8, 3],[1, 9],[9, 6],[10, 1],[10, 6],[10, 5]
Answer:
0 - 0
7 - 1
1 - 2
9 - 3
10 - 3
6 - 4
5 - 4
2 - 5
3 - 5
8 - 6
4 - 7

Input:
Graph G has 11 nodes, numbered from 0 to 10. Graph G has 17 edges.
This is the list of edge connections:
[0, 1],[1, 2],[3, 4],[4, 5],[5, 6],[0, 5],[3, 0],[3, 2],[0, 7],[7, 8],[1, 8],[9, 4],[7, 9],[2, 5],[10, 6],[10, 1],[10, 3]
Answer:
0 - 0
1 - 1
5 - 1
3 - 1
7 - 1
2 - 2
8 - 2
10 - 2
4 - 2
6 - 2
9 - 2

Input:
Graph G has 10 nodes, numbered from 0 to 9. Graph G has 10 edges.
This is the list of edge connections:
[5, 0],[6, 1],[6, 2],[2, 7],[7, 4],[8, 2],[9, 7],[9, 8],[9, 5],[3, 5]
Answer:
\end{gptquery}

\begin{gptanswer}
0 - 1
5 - 1
6 - 1
2 - 2
7 - 2
1 - 2
8 - 3
9 - 3
4 - 3
3 - 4
\end{gptanswer}

\section{Sorting nodes within layers}
\label{app:task-sorting-nodes}


\texttt{Sorting nodes - Standard:}
\begin{gptquery}
This is a list of directed edges in a layered graph.
[(2, 5), (1, 6), (6, 3), (0, 1), (2, 4), (2, 7), (1, 8), (1, 2), (1, 9), (9, 5), (5, 10), (7, 10)]

This is the description of what nodes are contained in what layer:  Layer 0: [0] 
Layer 1: [1] 
Layer 2: [6, 8, 2, 9] 
Layer 3: [3, 5, 4, 7] 
Layer 4: [10] 

Write no code and no explanations.

Start from layer 0. 
Consider every node in layer 0, and for each node consider its neighbors in the next layer. 
For each node in layer 0, calculate the positions of the neighbors in the next layer. 
Then, sort the nodes in layer 0 based on the median position of their neighbors.
If a number is NaN, then it is equal to 0.
If a node has no neighbors, assign an associated median of 0.

Repeat the process for every layer.

Finally, I want you to write the layers and the nodes in them once more, but the nodes in the layers have to be sorted according to their associated median value.
Like this:
Layer 0: [<sorted nodes in layer 0>]
If node A has an associated median value that is less than the associated median value of node B, then node A must come before node B. 
\end{gptquery}

\begin{gptanswer}
Layer 0: [0]
Layer 1: [1]
Layer 2: [8, 6, 2, 9]
Layer 3: [7, 5, 4, 3]
Layer 4: [10]
\end{gptanswer}

\texttt{Sorting nodes - Steps:}

\begin{gptquery}
This is a list of directed edges in a layered graph.
[(2, 5), (1, 6), (6, 3), (0, 1), (2, 4), (2, 7), (1, 8), (1, 2), (1, 9), (9, 5), (5, 10), (7, 10)]

This is the description of what nodes are contained in what layer:  Layer 0: [0] 
Layer 1: [1] 
Layer 2: [6, 8, 2, 9] 
Layer 3: [3, 5, 4, 7] 
Layer 4: [10] 

Start from layer 0. 
List every node in layer 0, and for each node list its neighbors in the next layer. 
Like this: 
Layer 0: [<node1>, <node2>, ...]
Neighbors of <node1>: [<neighbor1>, <neighbor2>, ...]

Then, for every one of the neighbors, write their index in their layer. So if node 6 belongs to layer 2 and is in the fourth position in layer 2, its index is 3.
Like this:
Index of <neighbor1>: <value>
And under each set of indices write the median of the indices.

Now I want you to start again from layer 0, and this time I want you to list the nodes in every layer and their associated medians.
Like this:
Layer 0:
<node 0> -> <median of the neighbors of node 0>

If a number is NaN, then it is equal to 0.
If a node has no neighbors, assign an associated median of 0.

Finally, I want you to write the layers and the nodes in them once more, but the nodes in the layers have to be sorted according to their associated median value. 
Like this:
Layer 0: [<sorted nodes in layer 0>]
If node A has an associated median value that is less than the associated median value of node B, then node A must come before node B. 
This time, write no additional information other than the layers and the sorted nodes.
\end{gptquery}

\begin{gptanswer}
Here are the requested outputs:

First output:

Layer 0: [0]
Neighbors of 0: []

Layer 1: [1]
Neighbors of 1: [6, 8, 2, 9]

Layer 2: [6, 8, 2, 9]
Neighbors of 6: [3]
Neighbors of 8: []
Neighbors of 2: [5, 4, 7]
Neighbors of 9: [5]

Layer 3: [3, 5, 4, 7]
Neighbors of 3: []
Neighbors of 5: [10]
Neighbors of 4: []
Neighbors of 7: [10]

Layer 4: [10]
Neighbors of 10: []

Indices:

Index of 6: 0
Index of 8: 1
Index of 2: 2
Index of 9: 3
Median of indices: 1.5

Index of 3: 0
Index of 5: 1
Index of 4: 2
Index of 7: 3
Median of indices: 1.5

Second output:

Layer 0:
0 -> 0

Layer 1:
1 -> 1.5

Layer 2:
6 -> 0
8 -> 0
2 -> 2
9 -> 3
Median of layer: 1.5

Layer 3:
3 -> 0
5 -> 1
4 -> 2
7 -> 3
Median of layer: 1.5

Layer 4:
10 -> 0

Third output:

Layer 0: [0]
Layer 1: [1]
Layer 2: [8, 6, 2, 9]
Layer 3: [5, 3, 4, 7]
Layer 4: [10]
\end{gptanswer}

\texttt{Sorting nodes - ICL:}

\begin{gptquery}
We want to reduce edge crossings on a layered graph drawing.
You are given a  list of directed edges in a layered graph, and a mapping of which nodes are contained in which layer.
Start from layer 0. 
Consider every node in layer 0, and for each node consider its neighbors in the next layer. 
For each node in layer 0, calculate the positions of the neighbors in the next layer. 
Then, sort the nodes in layer 0 based on the median position of their neighbors.
If a number is NaN, then it is equal to 0.
If a node has no neighbors, assign an associated median of 0.
Repeat the process for every layer.
Finally, I want you to write the layers and the nodes in them once more, but the nodes in the layers have to be sorted according to their associated median value.
Like this:
Layer 0: [<sorted nodes in layer 0>]
If node A has an associated median value that is less than the associated median value of node B, then node A must come before node B.

## Input:
Directed edges:
[(3, 2), (0, 3), (4, 5), (4, 1), (3, 4), (1, 7), (8, 7), (2, 8), (4, 6), (3, 9)]
Layer node mapping:
Layer 0: [0]
Layer 1: [3]
Layer 2: [2, 4, 9]
Layer 3: [8, 5, 1, 6]
Layer 4: [7]

## Answer:
Layer 0: [0]
Layer 1: [3]
Layer 2: [4, 9, 2]
Layer 3: [6, 5, 1, 8]
Layer 4: [7]

## Input:
Directed edges:
[(0, 1), (3, 2), (4, 3), (4, 5), (2, 7), (8, 6), (7, 8), (4, 9), (0, 4)]
Layer node mapping:
Layer 0: [0]
Layer 1: [1, 4]
Layer 2: [3, 5, 9]
Layer 3: [2]
Layer 4: [7]
Layer 5: [8]
Layer 6: [6]

## Answer:
Layer 0: [0]
Layer 1: [4, 1]
Layer 2: [9, 3, 5]
Layer 3: [2]
Layer 4: [7]
Layer 5: [8]
Layer 6: [6]

## Input:
Directed edges:
[(2, 5), (1, 4), (3, 1), (9, 2), (9, 3), (9, 8), (9, 6), (9, 7), (0, 9), (6, 1)]
Layer node mapping:
Layer 0: [0]
Layer 1: [9]
Layer 2: [2, 3, 8, 6, 7]
Layer 3: [5, 1]
Layer 4: [4]

## Answer:
Layer 0: [0]
Layer 1: [9]
Layer 2: [2, 7, 6, 3, 8]
Layer 3: [5, 1]
Layer 4: [4]

## Input:
Directed edges:
[(0, 8), (8, 1), (8, 2), (2, 7), (7, 3), (6, 3), (2, 6), (6, 4), (9, 5), (4, 9)]
Layer node mapping:
Layer 0: [0]
Layer 1: [8]
Layer 2: [1, 2]
Layer 3: [7, 6]
Layer 4: [3, 4]
Layer 5: [9]
Layer 6: [5]

## Answer:
Layer 0: [0]
Layer 1: [8]
Layer 2: [1, 2]
Layer 3: [6, 7]
Layer 4: [4, 3]
Layer 5: [9]
Layer 6: [5]

## Input:
Directed edges:
[(0, 5), (5, 1), (1, 6), (8, 4), (0, 8), (9, 2), (8, 9), (5, 10), (10, 2), (8, 10), (10, 6), (10, 7), (10, 3)]
Layer node mapping:
Layer 0: [0]
Layer 1: [5, 8]
Layer 2: [1, 10, 4, 9]
Layer 3: [6, 2, 7, 3]

## Answer:
Layer 0: [0]
Layer 1: [5, 8]
Layer 2: [1, 10, 4, 9]
Layer 3: [6, 3, 7, 2]

## Input:
Directed edges:
[(2, 5), (1, 6), (6, 3), (0, 1), (2, 4), (2, 7), (1, 8), (1, 2), (1, 9), (9, 5), (5, 10), (7, 10)]
Layer node mapping:
Layer 0: [0]
Layer 1: [1]
Layer 2: [6, 8, 2, 9]
Layer 3: [3, 5, 4, 7]
Layer 4: [10]

## Answer:
\end{gptquery}

\begin{gptanswer}
Layer 0: [0]
Layer 1: [1]
Layer 2: [8, 9, 2, 6]
Layer 3: [7, 5, 4, 3]
Layer 4: [10]
\end{gptanswer}

\section{Counting crossings}
\label{app:task-count-crossings}

This appendix provides more details for the counting crossings task (sometimes really identifying crossings) covered in
\iflabelexists{sec:task-count-cross-edge-length}
    {\cref{sec:task-count-cross-edge-length} of}
    {}
the paper.
These boxes show example wording for the \textbf{Query} to ChatGPT and the \textbf{Answer} we received.


\texttt{Counting crossings - Standard:}
\begin{gptquery}
Given the following arrays: 
A: [6, 4, 7, 8, 9, 2]
B: [5, 10]

And the following list of tuples:
Tuples: [(4, 5), (6, 5), (6, 10), (4, 10)]

Where for each tuple the first element of comes from array A, and the second element comes from array B.

Assuming this is a bipartite graph, count the edge crossings. Two edges that share a source or a target can not cross. Two edges cross if the order of their sources is opposite to the order of their targets.

Exclude all edge crossings where edges have the same source or the same target.

Write no explanations and no code. Return the pairs of edges that cross, one per line. There might be no edge crossings - in that case, return an empty list.

\end{gptquery}

\begin{gptanswer}
(6, 5) and (4, 10)
(6, 10) and (4, 5)
\end{gptanswer}

\texttt{Counting crossings - Steps:}

\begin{gptquery}
You are a very advanced program that can help me with counting edge crossings in a bipartite graph. Write the answer to the following question.

Given the following arrays: 
A: [6, 4, 7, 8, 9, 2]
B: [5, 10]

And the following list of edges:
Tuples: [(4, 5), (6, 5), (6, 10), (4, 10)]

Where for each edges the first element comes from array A, and the second element comes from array B.

VERY IMPORTANT: If there is a single edge, write 0 and stop writing!. If array A has length 1, write 0 and stop writing!. If array B has length 1, write 0 and stop writing!

Otherwise keep going and consider all the combinations of edge pairs. 

Then, exclude all the pairs that have the same first element. Write the list.
Then, exclude all the pairs that have the same second element. Write the list again.

Now for every pair of edges left, consider the first one. Say that s1 is the index of its first element in A, and t1 is the index of its second element in B. Then consider the second edge. Say that s2 is the index of its first element in A, and t2 is the index of its second element in B.
Write down s1, t1, s2 and t2 for every pair, like this:
<pair> => s1 = <value>, t1 = <value>, s2 = <value>, t2 = <value>

Now exclude all the pairs for which s1 > s2 and t1 < t2. Write the list again.

Now exclude all the pairs for which s1 < s2 and t1 > t2. Write the list again.

Write the number of edges left as a number.
\end{gptquery}

\begin{gptanswer}
First, let's exclude the cases where there is only one edge or one of the arrays has length 1, as instructed:

There is more than one edge, so we keep going.
Array A has more than one element, so we keep going.
Array B has more than one element, so we keep going.
Next, let's exclude the pairs that have the same first element and the pairs that have the same second element:

Pairs with the same first element: [(6, 5), (6, 10)].
Pairs with the same second element: [(4, 10)].
Now, let's consider every pair of edges left:

(4, 5) and (6, 5): (4,5) => s1 = 1, t1 = 0, s2 = 0, t2 = 0; (6,5) => s1 = 2, t1 = 0, s2 = 0, t2 = 0.
(4, 5) and (4, 10): (4,5) => s1 = 1, t1 = 0, s2 = 0, t2 = 1; (4,10) => s1 = 1, t1 = 0, s2 = 0, t2 = 1.
(6, 5) and (4, 10): (6,5) => s1 = 2, t1 = 0, s2 = 0, t2 = 1; (4,10) => s1 = 1, t1 = 0, s2 = 0, t2 = 1.
Next, let's exclude the pairs for which s1 > s2 and t1 < t2, and s1 < s2 and t1 > t2:

(4, 5) and (6, 5): s1 < s2 and t1 < t2, so we keep this pair.
(4, 5) and (4, 10): s1 = s2, so we exclude this pair.
(6, 5) and (4, 10): s1 > s2 and t1 > t2, so we exclude this pair.
Therefore, we are left with only one edge pair that satisfies all the conditions, which is (4, 5) and (6, 5). Hence, the number of edges left is 1.
\end{gptanswer}

\texttt{Counting crossings - ICL:}

\begin{gptquery}
You are a very advanced program that can help me with counting edge crossings in a bipartite graph.
I will provide you with the arrays of nodes of layers A and B and a list of edges as tuples.
For each edge the first element comes from array A, and the second element comes from array B.

Assuming this is a bipartite graph, count the edge crossings. 
Two edges that share a source or a target can not cross. 
Two edges cross if the order of their sources is opposite to the order of their targets.

Exclude all crossings where edges have the same source or the same target.

Write no explanations and no code. Return the number of edges that cross. There might be no edge crossings - in that case, return 0.

## Input:
Layer arrays:
A: [0]
B: [5]
Edge tuples:
Tuples: [(0, 5)]

## Answer:
0

## Input:
Layer arrays:
A: [1]
B: [6, 7]
Edge tuples:
Tuples: [(1, 6), (1, 7)]

## Answer:
0

## Input:
Layer arrays:
A: [7, 4, 2, 3]
B: [9, 8, 5]
Edge tuples:
Tuples: [(3, 8), (3, 9), (4, 9), (2, 8), (2, 5)]

## Answer:
3

## Input:
Layer arrays:
A: [9]
B: [5]
Edge tuples:
Tuples: [(9, 5)]

## Answer:
0

## Input:
Layer arrays:
A: [3, 5, 4, 7]
B: [10]
Edge tuples:
Tuples: [(5, 10), (7, 10)]

## Answer:
0

## Input:
Layer arrays:
A: [6, 4, 7, 8, 9, 2]
B: [5, 10]
Edge tuples:
Tuples: [(4, 5), (6, 5), (6, 10), (4, 10)]

## Answer:
\end{gptquery}

\begin{gptanswer}
2
\end{gptanswer}

\section{Edge length}
\label{app:task-edge-length}

This appendix supplements the discussion of the edge length task in
\iflabelexists{sec:task-count-cross-edge-length}
    {\cref{sec:task-count-cross-edge-length} of}
    {}
the paper.
The boxes below show the \textbf{Query} to ChatGPT and the \textbf{Answer} it gave us.


\texttt{Edge length - Standard:}
\begin{gptquery}
The following is the description of a layered graph. 
A layered graph is a graph where each node is contained in a single layer.
The following is the list of directed edges, formatted as [(<source_id>, <target_id>)].

edge_list: [(0, 6), (1, 6), (7, 4), (3, 8), (2, 9), (2, 10), (0, 10), (10, 4)]

The second is a mapping of layers to the nodes contained within. Every array is a different layer, and the numbers in every array indicate the nodes in that layer.

Layer 0: [0] 
Layer 1: [1, 2, 3, 7] 
Layer 2: [6] 
Layer 3: [10] 
Layer 4: [5, 8, 9] 
Layer 5: [4] 

Count the total edge length. The edge length of each edge e is always equal to the absolute value of the number of the layer the target is contained in, minus the number of the layer the source is contained in. The edge length can't be negative.

Write no explanations and no code. Return the total sum of the lengths.
\end{gptquery}

\begin{gptanswer}
30
\end{gptanswer}

\texttt{Edge length - Steps:}

\begin{gptquery}
The following is the description of a layered graph. 
A layered graph is a graph where each node is contained in a single layer.
The following is the list of directed edges, formatted as [(<source_id>, <target_id>)].

edge_list: [(0, 6), (1, 6), (7, 4), (3, 8), (2, 9), (2, 10), (0, 10), (10, 4)]

The second is a mapping of layers to the nodes contained within. Every array is a different layer, and the numbers in every array indicate the nodes in that layer.

Layer 0: [0] 
Layer 1: [1, 2, 3, 7] 
Layer 2: [6] 
Layer 3: [10] 
Layer 4: [5, 8, 9] 
Layer 5: [4] 

Count the total edge length. The edge length of each edge e is always equal to the absolute value of the number of the layer the target is contained in, minus the number of the layer the source is contained in. The edge length can't be negative.

First, for each edge, write the layer of the source node and the layer of the target node.
Then, write each edge and its associated length.
In the end, write the total sum of the lengths.
Write the final sum in a separate line at the end.
\end{gptquery}

\begin{gptanswer}
The layer of the source node and the layer of the target node for each edge are:

(0, 6) -> Layer 0 to Layer 2
(1, 6) -> Layer 1 to Layer 2
(7, 4) -> Layer 1 to Layer 5
(3, 8) -> Layer 1 to Layer 4
(2, 9) -> Layer 1 to Layer 4
(2, 10) -> Layer 1 to Layer 3
(0, 10) -> Layer 0 to Layer 3
(10, 4) -> Layer 3 to Layer 5

The associated length for each edge is:

(0, 6) -> 2 - 0 = 2
(1, 6) -> 2 - 1 = 1
(7, 4) -> 5 - 1 = 4
(3, 8) -> 4 - 1 = 3
(2, 9) -> 4 - 1 = 3
(2, 10) -> 3 - 1 = 2
(0, 10) -> 3 - 0 = 3
(10, 4) -> 5 - 3 = 2

The total sum of the lengths is:

2 + 1 + 4 + 3 + 3 + 2 + 3 + 2 = 18

Therefore, the final sum is 18.
\end{gptanswer}

\texttt{Edge length - ICL:}

\begin{gptquery}
A layered graph is a graph where each node is contained in a single layer.
A layered graph is described by list of directed edges, formatted as [(<source_id>, <target_id>)], and a mapping of layers to the nodes contained within.
Every array in the mapping is a different layer, and the numbers in every array indicate the nodes in that layer.

Count the total edge length. 
The edge length of each edge e is always equal to the absolute value of the number of the layer the target is contained in, minus the number of the layer the source is contained in.
The edge length can't be negative.

Write no explanations and no code. Return the total sum of the lengths.

## Input:
List of edges formatted as [<source_id>, <target_id>]:
edges = [(0, 1), (1, 2), (0, 3), (5, 4), (5, 6), (1, 6), (1, 8), (8, 2), (7, 4), (9, 2), (9, 3), (7, 10)]
Mapping of layers to nodes:
ranks = Layer 0: [0]
Layer 1: [1, 5, 7, 9]
Layer 2: [8]
Layer 3: [6, 10]
Layer 4: [3]
Layer 5: [2, 4]
## Answer:
34

## Input:
List of edges formatted as [<source_id>, <target_id>]:
edges = [(0, 1), (1, 2), (3, 2), (0, 3), (7, 6), (7, 1), (1, 6), (4, 1), (4, 3), (1, 8), (2, 8), (2, 5), (9, 5), (9, 6), (3, 10), (1, 10), (0, 10)]
Mapping of layers to nodes:
ranks = Layer 0: [0]
Layer 1: [4, 7, 9]
Layer 2: [3]
Layer 3: [1]
Layer 4: [2, 6, 10]
Layer 5: [5, 8]
## Answer:
35

## Input:
List of edges formatted as [<source_id>, <target_id>]:
edges = [(0, 7), (1, 7), (6, 2), (5, 2), (5, 3), (0, 2), (8, 9), (9, 3), (10, 6), (10, 9)]
Mapping of layers to nodes:
ranks = Layer 0: [0]
Layer 1: [1, 4, 5, 8]
Layer 2: [10]
Layer 3: [6, 9]
Layer 4: [2, 3, 7]
## Answer:
23

## Input:
List of edges formatted as [<source_id>, <target_id>]:
edges = [(0, 6), (6, 1), (7, 1), (7, 2), (7, 4), (9, 3), (4, 9), (5, 10)]
Mapping of layers to nodes:
ranks = Layer 0: [0]
Layer 1: [7]
Layer 2: [2, 4, 6]
Layer 3: [5, 9]
Layer 4: []
Layer 5: [1, 3, 8, 10]
## Answer:
16

## Input:
List of edges formatted as [<source_id>, <target_id>]:
edges = [(4, 2), (2, 5), (1, 5), (6, 1), (0, 7), (3, 8), (1, 2), (6, 9), (0, 9), (10, 5)]
Mapping of layers to nodes:
ranks = Layer 0: [0]
Layer 1: [3, 4, 6, 7, 10]
Layer 2: [8, 9]
Layer 3: [1]
Layer 4: [2]
Layer 5: [5]
## Answer:
18

## Input:
List of edges formatted as [<source_id>, <target_id>]:
edges = [(0, 6), (1, 6), (7, 4), (3, 8), (2, 9), (2, 10), (0, 10), (10, 4)]
Mapping of layers to nodes:
ranks = Layer 0: [0]
Layer 1: [1, 2, 3, 7]
Layer 2: [6]
Layer 3: [10]
Layer 4: [5, 8, 9]
Layer 5: [4]
## Answer:
\end{gptquery}

\begin{gptanswer}
15
\end{gptanswer}

\section{Poetry}
\label{app:task-poetry}

Does ChatGPT know the Sugiyama layout algorithm? We asked it to explain it---in poetry!

\begin{gptquery}
Explain the Sugiyama layout algorithm, in poetry
\end{gptquery}
\begin{gptanswer}
In the land of graphs and nodes,
Where edges connect in different modes,
There lies a layout algorithm called Sugiyama,
That arranges graphs with utmost drama.

First, it separates the graph into layers,
Like floors in a building with multiple players,
Then assigns a rank to each node,
Determining which layer they call their abode.

Next, it creates a virtual layout,
That sets the nodes in their rightful state,
A rank and a position, left or right,
It lays them out with careful might.

Then, it adds dummy nodes in between,
To make sure edges are seen,
They help the algorithm avoid collisions,
And ensure graph beauty with precision.

Finally, it applies a few more rules,
To make sure the layout's cool,
Like minimizing edge lengths and crossings,
And smoothing out the lines without lossings.

And so, the Sugiyama algorithm is done,
A layout masterpiece that's second to none,
It turns messy graphs into organized art,
And helps us see the connections with a new start.
\end{gptanswer}

\section{Graph generation}
\label{app:task-graph-generation}

Generating random graphs can be fundamental for testing the results of a layout algorithm.
Authors of layout algorithms might be interested in graphs with particular features to test edge cases: for instance, a particular distribution of node degrees (either sparse, or highly connected) can help test the performance of an algorithm under such conditions. 

\begin{gptquery}
Generate a graph for me. Write no explanations, exclusively write the json representation of the graph. I want the graph to have 5 nodes and 7 edges.
\end{gptquery}

ChatGPT gave us valid and useful answers, e.g., for the above query it returned:

\begin{gptanswer}
{
  "nodes": [
    {"id": 1, "label": "Node 1"},
    {"id": 2, "label": "Node 2"},
    {"id": 3, "label": "Node 3"},
    {"id": 4, "label": "Node 4"},
    {"id": 5, "label": "Node 5"},
  ],
  "edges": [
    {"source": 1, "target": 2},
    {"source": 1, "target": 3},
    {"source": 1, "target": 4},
    {"source": 2, "target": 3},
    {"source": 3, "target": 4},
    {"source": 4, "target": 5},
    {"source": 5, "target": 1}
  ]
}
\end{gptanswer}

We also tested the ability of ChatGPT to define graphs that include multivariate data, while describing the kind of data we needed associated to nodes and edges. This opens up the possibility for ChatGPT to be used to generate temporal and weighted graphs.
\begin{gptquery}
Generate a graph for me. Write no explanations, exclusively write the json representation of the graph. I want the graph to have 5 nodes and 7 edges. I want every node to have a timestamp from January 1, 1970 to December 31, 1970. I want every edge to have a weight associated with it, which must be higher than 2.
\end{gptquery}

Again, ChatGPT delivered a valid and useful answer:

\begin{gptanswer}
{
  "nodes": [
    { "id": 1, "timestamp": "1970-01-01" },
    { "id": 2, "timestamp": "1970-02-01" },
    { "id": 3, "timestamp": "1970-03-01" },
    { "id": 4, "timestamp": "1970-04-01" },
    { "id": 5, "timestamp": "1970-05-01" }
  ],
  "edges": [
    { "source": 1, "target": 2, "weight": 3 },
    { "source": 1, "target": 3, "weight": 4 },
    { "source": 1, "target": 4, "weight": 5 },
    { "source": 2, "target": 3, "weight": 6 },
    { "source": 2, "target": 4, "weight": 7 },
    { "source": 3, "target": 4, "weight": 8 },
    { "source": 4, "target": 5, "weight": 9 }
  ]
}
\end{gptanswer}

\section{Format conversion}
\label{app:task-format-conversion}

Changing between different graph representation formats can be a tedious job that takes time away from more involved tasks---from time to time, we might need to transform a dataset from a CSV format to a JSON one, or various other formats that might be needed for the software we are using.
As our experiments showed that GPT had no problem converting a common graph format to another, we decided to use it as a pre-processing step in our own sample application of a layout algorithm.

The original format of the Rome-Lib graphs we used in the experiments is \texttt{graphml}, an XML-like format which is unnecessarily verbose for the purpose we wanted to use it for and a relatively less common graph format.
Since we wanted to format the graph to be suitable for a ChatGPT query, we needed to change the format into a less verbose one with fewer unnecessary tokens in the input that could limit the length of the response.
We opted---as a way to concisely describe a graph---to use a list of edges formatted as \texttt{[<source node id>, <target node id>]}, so that the input for ChatGPT looked like:

\begin{gptquery}
Graph G has 5 nodes, numbered from 1 to 5.
Graph G has 4 edges.
  
This is the list of edge connections:
[1, 2]
[2, 3]
[2, 4]
[2, 5]
\end{gptquery}

Instead of writing our own code or using an existing library to parse the \texttt{graphml} file and transform it into our more concise format, we asked ChatGPT to write the code to do that for us!
Refining the answer to obtain the code with the right functionality took us approximately 30 minutes.

\section{Defining new properties}
\label{app:task-defining-properties}

Although some properties of interest may already be known by ChatGPT (e.g.\ planarity), we are interested in seeing how ChatGPT behaves with newly-defined properties.
Can a user introduce properties, metrics, or other aspects of the graph on the fly?

To test this, we defined two new properties that have no predefined meaning in the context of graphs: ``bulbaceous'' and ``flamboyous''.
ChatGPT proved it could correctly understand how to apply these new properties.

Even though ``bulbaceous'' is a very simple property for which no knowledge of graph structure is needed, the purpose of this test was to see if ChatGPT could perform this task within the context of graphs:
\begin{gptquery}
A graph is bulbaceous if its number of edges is a multiple of its number of nodes. 
 
Graph G has 2 nodes and 4 edges.
 
Is graph G bulbaceous?
\end{gptquery}

\begin{gptanswer}
Yes, graph G is bulbaceous. The definition of a bulbaceous graph states that the number of edges must be a multiple of the number of nodes, and in the case of graph G, the number of edges (4) is a multiple of the number of nodes (2). Therefore, graph G is bulbaceous.
\end{gptanswer}

\begin{gptquery}
A graph is bulbaceous if its number of edges is a multiple of its number of nodes. 
 
Graph G has 2 nodes and 6 edges.
 
Is graph G bulbaceous?

Write no explanations.
\end{gptquery}

\begin{gptanswer}
Yes.
\end{gptanswer}

The ``flamboyous'' property is a little more complicated:
\begin{gptquery}
A graph is flamboyous if there are no two edges connecting the same set of nodes. 
 
Graph G has 3 nodes (0, 1 and 2) and 5 edges.
The edges of G are: (0, 1) (0, 1) (0, 2) (1, 2) (1, 2)
 
Is graph G flamboyous?
 
Write no explanations.
\end{gptquery}

\begin{gptanswer}
No.
\end{gptanswer}

\begin{gptquery}
A graph is flamboyous if there are no two edges connecting the same set of nodes. 
 
Graph G has 4 nodes (0, 1 , 2 and 3) and 4 edges.
The edges of G are: (0, 1) (1, 2) (2, 3) (3, 4)
 
Is graph G flamboyous?
 
Write no explanations.
\end{gptquery}

\begin{gptanswer}
The answer is: Yes.
\end{gptanswer}

\begin{figure}[h]
    \centering
    \includegraphics[width=.6\linewidth]{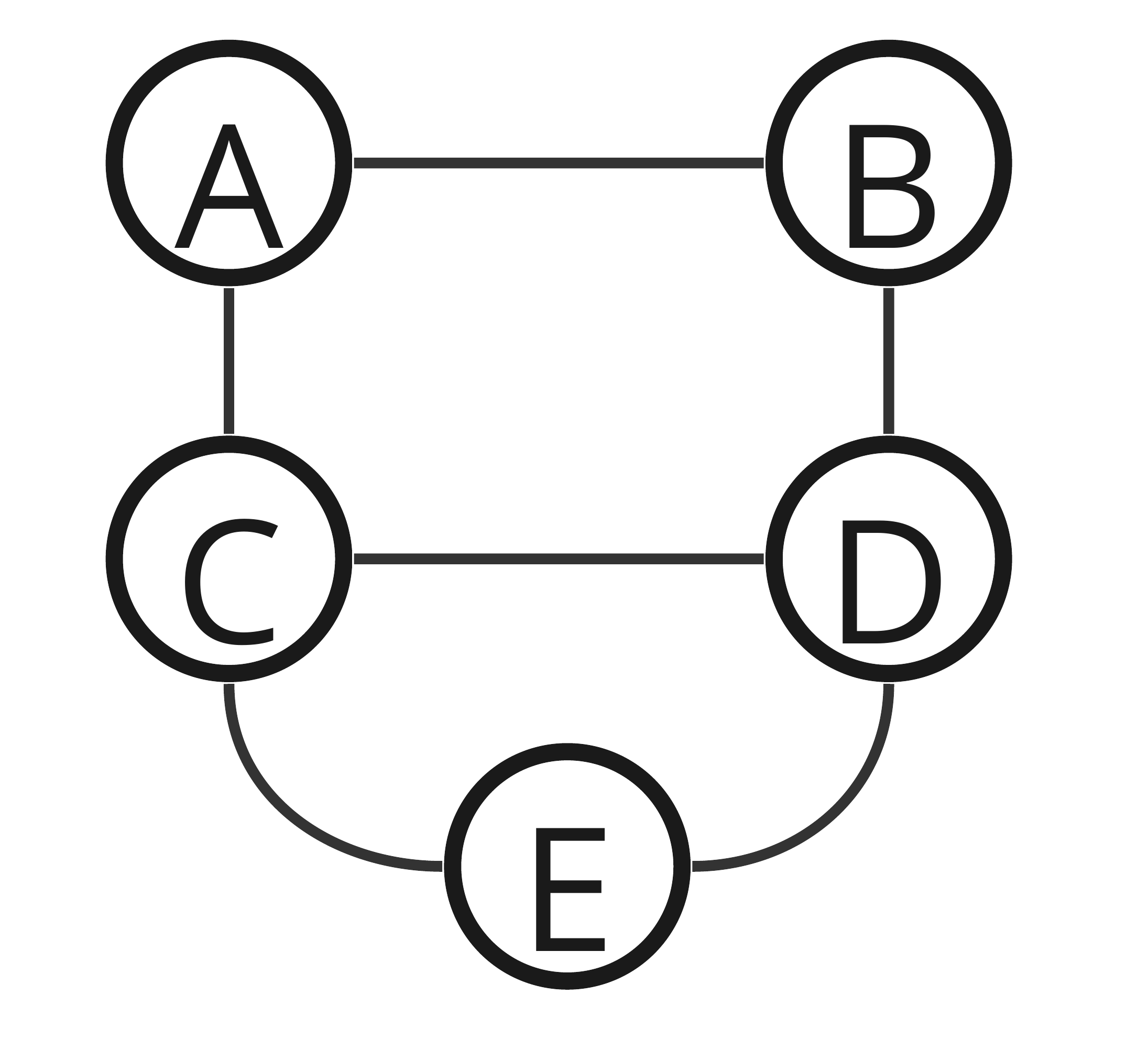}\\
    \includegraphics[width=.26\linewidth]{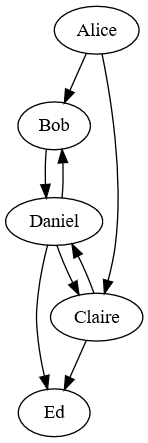}
    \includegraphics[width=.33\linewidth]{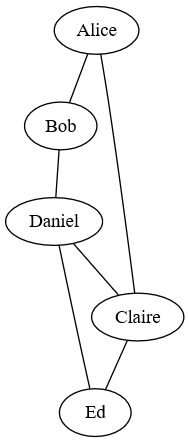}
    \includegraphics[width=.29\linewidth]{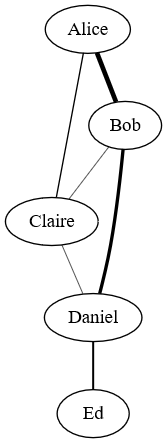}
    \caption{Simple topology used as ground truth for the written scene (top).
    The bottom shows the directed, non-weighted, and weighted answers provided by ChatGPT, respectively.}
    \label{fig:scenegraph}
\end{figure}

\section{Generating a graph from a scene}
\label{app:generate-graph-from-scene}
Creating a graph from textual information can be a challenging or cumbersome task, because the data might not be structured in a logical graph structure.
In our idealized scenario, a domain expert should be able to generate a graph from varied representations to achieve their goals, for instance, from a textual description of a scene. 

For this experiment, we first drew a simple graph with 5 nodes and 6 edges (\Cref{fig:scenegraph}, top) to serve as a ground truth.
Then, we wrote a short paragraph where interactions between office colleagues corresponded to the ground truth.
Using the DOT graph format is a good way to quickly test results, e.g., using Graphviz, and give ChatGPT an expressive output. 

\begin{gptquery}
I am going to give you a description of a short scene, and I would like you to provide me with a DOT graph representation in a code block where each person in the scene is a node, and whenever two characters interact there is an edge between them:

When first entering the office, Alice greeted Bob who was sitting at his desk and working at the final report. Until the end of the day, Alice never left her cubicle, only briefly talking with Claire during lunch. Most of the afternoon, Daniel and Bob were doing pair programming in the side room, and at the end of the day Dani, Claire and Ed had a meeting.
\end{gptquery}

\begin{gptanswer}
    digraph scene {
    Alice;
    Bob;
    Claire;
    Daniel;
    Ed;

    Alice -> Bob;
    Alice -> Claire;
    Bob -> Daniel;
    Daniel -> Bob;
    Daniel -> Claire;
    Daniel -> Ed;
    Claire -> Daniel;
    Claire -> Ed;
}
The above DOT graph represents the scene described in the prompt. Each character is a node, and whenever two characters interact, there is an edge between them. For example, Alice greets Bob when she first enters the office, so there is an edge between Alice and Bob. Similarly, Alice briefly talks with Claire during lunch, so there is an edge between Alice and Claire. Daniel and Bob do pair programming, so there is an edge between them. Finally, Daniel, Claire, and Ed have a meeting at the end of the day, so there are edges between Daniel and Claire, Daniel and Ed, and Claire and Ed.\end{gptanswer}

\begin{gptquery}
    How are you choosing the directions of the edges?
\end{gptquery}
\begin{gptanswer}
    I am choosing the directions of the edges based on the order of the events in the prompt. For example, if the prompt says that Alice greeted Bob, then the edge would go from Alice to Bob, indicating that Alice initiated the interaction with Bob. Similarly, if the prompt says that Daniel presented his results to Claire and Ed, then the edge would go from Daniel to Claire and from Daniel to Ed, indicating that Daniel was the one presenting his results to Claire and Ed.
\end{gptanswer}

From this answer we can see that ChatGPT extrapolates and adds its own interpretation to the task, which is not necessarily bad.
The bottom-left of \Cref{fig:scenegraph} shows the resulting directed graph, which has ChatGPT's arbitrarily-assigned directions.
The topology (if turned into an undirected graph) is equivalent to the ground-truth!
When asked to give the equivalent, non-directed version, ChatGPT provided the bottom-middle graph, which is also correct. 

A final experiment, we decided to test ChatGPT's capacity to quantify these interactions between people as edge weights, similar to our idealized example from the Introduction.
The topology remained the same, but we added different degrees of interaction between each person in the query:
\begin{gptquery}
I am then giving you another variation of the scene, and I want you to quantify the amount of interactions between people with edge thickness, please:

When first entering the office, Alice greeted Bob who was sitting at his desk and working at the final report. Until the end of the day, Alice never left her cubicle, only briefly talking with Claire during lunch and sharing half of her sandwich with Bob. Most of the afternoon, Daniel and Bob were doing pair programming in the side room, and at the end of the day Dani, Claire and Ed had a meeting. After the meeting, Daniel and Ed stayed a little bit late working on a side project. Because Alice had given him half of her sandwich at lunch, Bob offered to give her a ride home and she accepted.
\end{gptquery}
\begin{gptanswer}
    graph scene {
    Alice -- Bob [penwidth=3.5];
    Alice -- Claire [penwidth=1.0];
    Bob -- Claire [penwidth=0.5];
    Bob -- Daniel [penwidth=2.5];
    Claire -- Daniel [penwidth=0.5];
    Daniel -- Ed [penwidth=1.5];
}

The above DOT graph represents the scene described in the prompt as an undirected graph, with the thickness of the edges quantifying the amount of interaction between the characters. The thicker the edge, the more interaction there is between the two characters.

For example, Alice greets Bob when she first enters the office, so there is a thick edge between Alice and Bob. Alice shares half of her sandwich with Bob, which is not represented in this graph. Bob offers to give Alice a ride home, which is not represented in this graph.

Similarly, Daniel and Bob do pair programming, so there is a thick edge between them. Daniel, Claire, and Ed have a meeting at the end of the day, so there are thinner edges between them. Finally, Daniel and Ed work on a side project, which is indicated by a thin edge between them.

Note that the thickness of the edges is somewhat subjective and can be adjusted based on the level of interaction that you want to emphasize. Also, the edges between Bob and Claire, and between Claire and Daniel, are very thin, indicating a small amount of interaction.

\end{gptanswer}

ChatGPT's answer (\Cref{fig:scenegraph}, bottom-right) strangely removes the edge between Claire and Ed, so we asked it to explain its reasoning:

\begin{gptquery}
can you present a more detailed explanation of your choices of edge thickness based on the interactions?
\end{gptquery}
\begin{gptanswer}
Certainly! Here is a more detailed explanation of how I assigned the thickness of the edges in the graph, based on the interactions described in the prompt:

    - Alice and Bob: Alice greets Bob when she first enters the office, and they briefly talk during lunch, with Alice sharing half of her sandwich with Bob. Bob also offers to give Alice a ride home. These interactions suggest a relatively strong connection between Alice and Bob, so I assigned a relatively thick edge (penwidth=3.5) between them.

    - Alice and Claire: Alice only briefly talks with Claire during lunch, which suggests a weaker connection between them. Therefore, I assigned a thinner edge (penwidth=1.0) between Alice and Claire.

    - Bob and Claire: Bob and Claire only interact briefly when Alice shares her sandwich with Bob during lunch, which suggests a very weak connection between them. Therefore, I assigned a very thin edge (penwidth=0.5) between Bob and Claire.

    - Bob and Daniel: Bob and Daniel spend most of the afternoon doing pair programming, which suggests a relatively strong connection between them. Therefore, I assigned a relatively thick edge (penwidth=2.5) between Bob and Daniel.

    Claire and Daniel: Claire and Daniel are both in the meeting at the end of the day, which suggests a weak connection between them. Therefore, I assigned a thin edge (penwidth=0.5) between Claire and Daniel.

    Daniel and Ed: Daniel and Ed are both in the meeting at the end of the day, and they also work on a side project together after the meeting. These interactions suggest a relatively strong connection between them, so I assigned a moderately thick edge (penwidth=1.5) between Daniel and Ed.

Note that the thickness of the edges is somewhat subjective and can be adjusted based on the level of interaction that you want to emphasize.
\end{gptanswer}
As ChatGPT itself said, this is a somewhat subjective decision, and ChatGPT is really using subtleties in language to come out with its interpretation.
This demonstrates how sensitive these models are to the choice of words, something we already observed in other experiments but not as explicitly as in this case. 
The results of this simple test case show promise in the direction of natural language interfaces for graph drawing.
The interfaces currently are not necessarily impressive, but ChatGPT's ability to give correct results and explain them is already very developed.

\section{Generating a scene from a graph}
\label{app:generate-scene-from-graph}

As a follow-up to the previous experiment, we decided to test the opposite direction: given a graph, can ChatGPT write an office scene with equivalent interactions for its edges?

\begin{gptquery}
I am going to give you a DOT graph format where each node is a person, and each edge represents that these people interacted in some way, such as chatting, having a meeting, working together, or something. I want you to write a short description of a day at the office that reproduces the topology of the graph:

graph scene {
    Alice;
    Bob;
    Claire;
    Daniel;
    Ed;

    Alice -- Bob;
    Alice -- Claire;
    Bob -- Daniel;
    Daniel -- Claire;
    Daniel -- Ed;
    Claire -- Ed;
}
\end{gptquery}
\begin{gptanswer}
It's a typical workday at the office, and the team is collaborating on a project. Alice is busy typing away at her computer, exchanging ideas with Bob and Claire, who are both sitting at their desks nearby. Bob and Alice catch up on some tasks they need to complete together, while Claire provides some helpful insights based on her experience with similar projects.

Meanwhile, Daniel is having a one-on-one meeting with Bob, discussing some new features they need to implement. They brainstorm ideas, and Daniel takes notes on his laptop. After their meeting, Bob heads over to Daniel's desk to get some more details on a specific task.

Later in the day, Claire and Daniel meet up to work on a presentation they need to deliver the following day. They huddle around Claire's computer, discussing the best way to convey their ideas to the audience. As they work, Ed pops over to Claire's desk to ask her about a deadline for a different project.

The day winds down, and the team members start to pack up their things. They wave goodbye and head out, ready to tackle another day of work together tomorrow.
\end{gptanswer}
This is an OK response, but we noticed that there was no explicit interaction between Ed and Daniel.
We asked ChatGPT where it was:
\begin{gptquery}
Which part of the scene you just wrote is representing or accounting for the edge between Ed and Daniel in the input graph?
\end{gptquery}

\begin{gptanswer}
The part of the scene that represents the edge between Ed and Daniel in the input graph is when Daniel is having a one-on-one meeting with Bob, and after the meeting, Bob heads over to Daniel's desk to get some more details on a specific task. At this point, Ed pops over to Claire's desk to ask her about a deadline for a different project, which implies that Ed and Daniel did not have a direct interaction during this particular workday. However, the edge between Ed and Daniel suggests that they have interacted in some way before or might do so in the future.
\end{gptanswer}
A bit of a cheeky answer, technically correct but once again showing how ChatGPT powerfully uses language to create subjective or interpretative responses.
The potential for generating data based on an input graph topology could have interesting applications, beyond just providing another modality of interacting with ChatGPT.
The user could draw the graph in any software representing some desired topology for a task, then turn it into a textual representation such as DOT used here to pass as a prompt to ChatGPT.

\section{Generating SVG representations of graphs}
\label{app:drawing}

\begin{figure*}[tb]
    \centering
    \includegraphics[width=.9\linewidth]{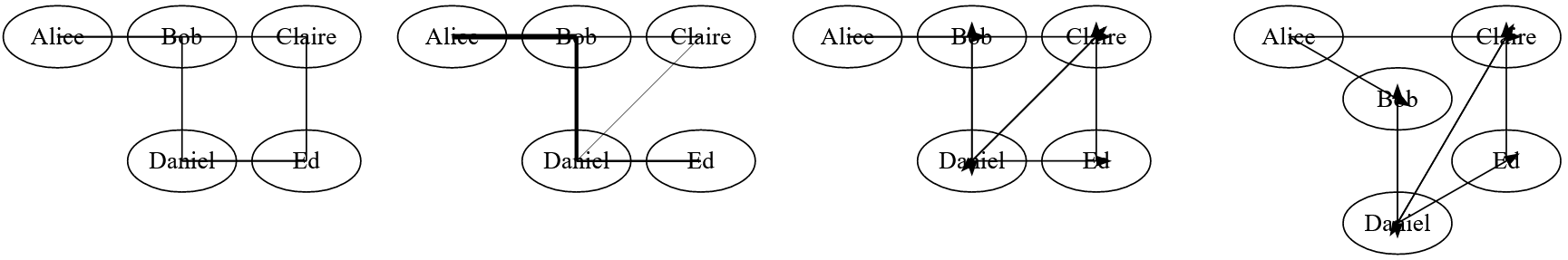}
    \caption{SVG outputs obtained by inputting the DOT graphs that ChatGPT generated from our scene prompts (\cref{app:generate-graph-from-scene}).
    From left to right, non-directed version, weighted version, directed version, and directed version after asking for a correction on node positions.}
    \label{fig:dottosvg}
\end{figure*}

The final step in a layout algorithm is usually to draw the graph using some rendering library.
We experimented with several formats and methods for rendering graphs using ChatGPT, including asking for scripts in different languages such as p5.js as well as asking for ASCII representations. 
The results were generally poor, with missing edges and nodes and outputs that disregarded the input graph.
A full report of this kind of experiment likely requires another paper.
The most straightforward way to get a drawing was to ask ChatGPT for a SVG representation.
We found that asking for a translation between DOT and SVG produced some interesting answers.

\Cref{fig:dottosvg} shows the result of the SVGs generated by providing ChatGPT with the DOT representations it generated from our textual scenes (\cref{app:generate-graph-from-scene}).
We refrain from pasting the full answers here because the returned SVGs are many lines of text.
Shown left-to-right, we first we asked for the non-directed version, then the weighted version, and the directed version. 
The fourth representation was the result of asking ChatGPT to better position the nodes because Alice, Bob, and Claire were on the same line.

Before obtaining the third version (directed) we were not sure if ChatGPT was respecting the topology, but the arrows made us realize that the three top nodes were aligned.
This resulted in the Alice--Claire edge appearing like two edges: Alice--Bob and Bob--Claire.
This is a terrible faux-pas by graph drawing standards, but the results are still interesting.
ChatGPT was able to successfully translate many features from the DOT representation into SVG and fix the layout when requested.

\end{document}